# Magnetic hysteresis in nanostructures with thermally-controlled RKKY coupling

1. Dmytro Polishchuk[1,2]
   [1]Nanostructure Physics, Royal Institute of Technology, 10691 Stockholm, Sweden
   [2]Institute of Magnetism, NAS of Ukraine and MES of Ukraine, 03142 Kyiv, Ukraine
   E-mail: dmytropol@gmail.com

2. Yuliya Tykhonenko-Polishchuk[1,2]
   [1]Nanostructure Physics, Royal Institute of Technology, 10691 Stockholm, Sweden
   [2]Institute of Magnetism, NAS of Ukraine and MES of Ukraine, 03142 Kyiv, Ukraine
   E-mail: nikaapterous@gmail.com

3. Vladyslav Borynskyi[2]
   [2]Institute of Magnetism, NAS of Ukraine and MES of Ukraine, 03142 Kyiv, Ukraine
   E-mail: vladislav.borinskiy@gmail.com

4. <u>Anatolii Kravets[1,2]</u>
   [1]Nanostructure Physics, Royal Institute of Technology, 10691 Stockholm, Sweden
   [2]Institute of Magnetism, NAS of Ukraine and MES of Ukraine, 03142 Kyiv, Ukraine
   <u>Corresponding author</u>
   E-mail: anatolii@kth.se

5. Alexandr Tovstolytkin[2],
   [2]Institute of Magnetism, NAS of Ukraine and MES of Ukraine, 03142 Kyiv, Ukraine
   E-mail: atov@imag.kiev.ua

6. Vladislav Korenivski[1]
   [1]Nanostructure Physics, Royal Institute of Technology, 10691 Stockholm, Sweden
   E-mail: vk@kth.se

**Abstract**

Mechanisms of the recently demonstrated ex-situ thermal control of the indirect exchange coupling in magnetic multilayer are discussed for different designs of the spacer layer. Temperature-induced changes in the hysteresis of magnetization are shown to be associated with different types of competing interlayer exchange interactions. Theoretical analysis indicates that the measured step-like shape and hysteresis of the magnetization loops is due to local in-plane magnetic anisotropy of nano-crystallites within the strongly ferromagnetic films. Comparison of the experiment and theory is used to contrast the mechanisms of the magnetization switching based on the competition of (i) indirect (RKKY) and direct (non-RKKY) interlayer exchange interactions as well as (ii) indirect ferromagnetic and indirect antiferromagnetic (both of RKKY type) interlayer exchange. These results, detailing the rich magnetic phase space of the system, should help enable the practical use of RKKY for thermally switching the magnetization in magnetic multilayers.

**Background**

The important discoveries of the indirect exchange coupling (IEC) [1] of Ruderman-Kittel-Kasuya-Yosida (RKKY) type and of the giant magnetoresistance effect [2] have generated a great deal of new basic physics results as well as numerous applications [3]. The discovered IEC oscillates in magnitude and sign versus the separation of the individual ferromagnetic layers in a metallic stack, yielding either parallel (P) or antiparallel (AP) magnetic ground states. This RKKY-type interaction is almost independent of temperature [4, 5] and largely insensitive to any other external control post-fabrication, which limits the use of the effect. Recent attempts to enhance the effect of temperature on RKKY and use it to control the IEC in Tb/Y/Gd [6] and Co/Pt [7] multilayers report relatively weak RKKY without direct parallel-to-antiparallel (P-to-AP) thermal switching, with broad thermal transitions (~100 K).

We recently demonstrated [8, *9*] a new mechanism of ex-situ thermal control of the interlayer RKKY coupling in magnetic multilayers. The idea is based on the use of a diluted ferromagnetic alloy with relatively low Curie temperature ($T_C^*$) instead of the nonmagnetic spacer between strongly ferromagnetic (FM) layers. In the initial design, the Cr spacer in a classical RKKY trilayer Fe/Cr/Fe is replaced with the diluted alloy $Fe_xCr_{100-x}$ [Fig. 1(a)–(b)]. When the spacer is paramagnetic (PM) at $T > T_C^*$ [Fig. 1(a)], the trilayers exhibit an antiparallel alignment of the Fe moments due to the antiferromagnetic (AFM) *indirect* exchange coupling (RKKY). The parallel alignment is enforced by the *direct* exchange coupling when the spacer is FM ($T < T_C^*$) [Fig. 1(b)]. As temperature is varied, these trilayers demonstrate a parallel-to-antiparallel magnetization switching, with a rather broad transition of ~100 K due to the magnetic proximity effect [10]. In contrast to the trilayers with the *uniform* spacer, trilayers with non-uniform, *composite* spacers demonstrate a significantly enhanced performance with the thermo-magnetic transition widths down to ~10 K. Moreover, by tailoring the spacer properties, either an antiparallel [Fig. 1(c)] or parallel ground state [Fig. 1(e)] can be obtained at $T < T_C^*$. On heating above $T_C^*$, the Fe moments reverse their mutual orientation either into parallel for Fe/sp1/Fe [Fig. 1(d)] or into antiparallel for Fe/sp2/Fe [Fig. 1(f)]. The narrow thermal transition and the ability to choose the magnetic regime (P/AP) as well as the operating temperature interval are all important advantages in terms of practical implications.

Antiferromagnetic interlayer coupling in our multilayers is clearly manifested as zero remnant magnetization, reflecting the antiparallel alignment of the Fe layers' moments. Besides the zero remanence, the magnetization curves $M(H)$ are characterized by a step-like approach to saturation and hysteresis on reversing the field sweep [Fig. 2(a)]. The bi-linear exchange coupling model gives $M(H)$ as a line approaching saturation at the effective field of the indirect exchange, $H_J$. A step-like character of the saturation occurs due to in-plane magnetic anisotropy of the structure's ferromagnetic layers, resulting in one step for the easy-axis anisotropy [11], and two sequential steps for the four-fold anisotropy [12]. However, our angle-dependent magnetometric and magnetoresonance studies of the multilayers reveal no macroscopic in-plane magnetic anisotropy. The latter fact warrants a more comprehensive analysis of the experimental results, accompanied by model simulations. In the following, such comprehensive approach is used to contrast the mechanisms of the magnetization switching for the two key multilayer designs – with uniform [Fig. 1(a),(b)] as against composite spacer layers [Fig. 1(c),(e)].

We point out the importance of understanding the mechanisms involved in the interlayer exchange in a given system. The pioneering work on RKKY in multilayers [13] and its extensions to, e.g., bi-quadratic exchange [14, 15] set off a major development in physics and technology known as spintronics. The RKKY in the original form, however, is not used today due to the lack of a suitable switching mechanism, but often plays an assisting role in devices for, e.g., flux-closing reference layers. In this work, we study such a primary RKKY-switching mechanism and, more specifically, analyze the interplay among the interactions leading to thermal on/off switching of RKYY, which in turn controls the efficiency of the P/AP switching of the magnetization of the nanostructure. Based on this analysis, we are able to make conclusions about and recommendations for optimizing the switching performance of the Curie-RKKY nanodevices.

**Methods**

In this work we analyze two series of samples: (1) Fe(2)/sp1($x$ = 30–40 at.%)/Fe(2), where sp1 = N/f/N/f/N, N = Cr(1.5), f = Fe(0.25)/Fe$_x$Cr$_{100-x}$(3)/Fe(0.25) [Fig. 1(c)], and (2) Fe(2)/sp2($x$ = 10–20 at.%)/Fe(2), where sp2 = N/f/N, N = Cr($d_{Cr}$), f = Fe$_x$Cr$_{100-x}$($d$), $d_{tot}$ = ($2d_{Cr} + d$) = 1.5 nm [Fig. 1(e)]. Additionally, a number of reference films and bi-layers were deposited. The thicknesses in parentheses are in "nm". The multilayers were deposited at room temperature onto Ar pre-etched undoped Si (100) substrates using a dc magnetron sputtering system. Layers of diluted Fe$_x$Cr$_{100-x}$ binary alloys of varied composition were deposited using co-sputtering from separate Fe and Cr targets. Additional details on the multilayer fabrication can be found elsewhere [8, 9].

The in-plane magnetic characterization was carried out using a vibrating-sample magnetometer (VSM) equipped with a high-temperature furnace (Lakeshore Inc.) in the temperature range of 295–400 K, and a magneto-optical Kerr effect (MOKE) magnetometer equipped with an optical cryostat (Oxford Instr.) in the temperature range of 77–450 K. Additionally, ferromagnetic resonance (FMR) measurements were performed room-temperature using an X-band Bruker ELEXYS E500 spectrometer equipped with an automatic goniometer to measure the in-plane-angle dependence of the magnetic resonance spectra.

**Results and discussion**

**Phenomenology of indirect exchange coupling**

A phenomenological magnetostatic model used for simulations of magnetization curves for trilayer F1/NM/F2, where F1 and F2 are ferromagnetic layers and NM is nonmagnetic spacer, has the following assumptions. First, the magnetic field is applied in the plane of the films, which corresponds to our experiment and simplifies the calculations. Second, the individual grains in the polycrystalline films are characterized by two-fold in-plane anisotropy with the easy axes uniformly distributed across all in-plane angles (the films were deposited under in-plane rotation). These assumptions are reasonable for the studied system and produced the best fit to the measured $M(H)$ data at various temperatures as discussed below.

The free energy density for our F1/NM/F2 system can then be written as

$$U = U_H + U_a + U_J =$$
$$= -MH\left[\cos(\varphi_1 - \varphi_H) + \cos(\varphi_2 - \varphi_H)\right] - \left(1/2\, MH_{a1}\cos^2\varphi_1 + 1/2\, MH_{a2}\cos^2\varphi_2\right) + \quad (1)$$
$$+ 1/2\, MH_J \cos(\varphi_1 - \varphi_2),$$

where $U_H$, $U_a$ and $U_J$ are, respectively, the Zeeman energy of the FM layers in field $\mathbf{H} = (H, \varphi_H)$, uniaxial anisotropy energy, and the interlayer coupling energy of bi-linear type [16, 17]. The magnetic moments of the FM layers, $\mathbf{M}_1 = (M, \varphi_1)$ and $\mathbf{M}_2 = (M, \varphi_2)$, are of the same magnitude, as illustrated in Fig. 2(c). $H_{a1,2}$ and $H_J$ are the effective fields of the uniaxial (two-fold) anisotropy and the bi-linear interlayer coupling, respectively. Conversion to angular variables $\varphi_m = (\varphi_1 + \varphi_2)/2$ and $\varphi_d = (\varphi_1 - \varphi_2)$ simplifies the expression for the magnetic free energy of the system to

$$U = -2MH\cos(\varphi_m - \varphi_H)\cos(\varphi_d/2) - 1/2\, M\left[H_{a1}\cos^2(\varphi_m + \delta/2)\right.$$
$$\left. + H_{a2}\cos^2(\varphi_m - \delta/2)\right] + 1/2\, MH_J \cos\varphi_d. \quad (2)$$

In the following simulations, the magnetization curves, $M(H)$, are obtained by finding parameters $\varphi_m$ and $\varphi_d$, which correspond to the minimum of $U$ in (2) for given $\varphi_H$, $H_{1a}$, $H_{2a}$, and $H_J$, according to

$$M/M_s = \left[\cos(\varphi_1 - \varphi_H) + \cos(\varphi_2 - \varphi_H)\right]/2 = \cos(\varphi_m - \varphi_H)\cos(\varphi_d/2). \quad (3)$$

**Coercivity of magnetization**

The measured $M(H)$ for the structures with AFM exchange coupling are of a step-like shape, with well-defined coercivity for the reversing field sweep [Fig. 3(a)]. The above phenomenological model is used to analyze both the magnetic properties of the ferromagnetic layers Fe(2 nm) and the thermally-induced magnetic transition in the composite spacers, which mediates the interlayer coupling.

Epitaxial (100) Fe based multilayers grown on single-crystal substrates are usually characterized by four-fold in-plane magnetic anisotropy [12], while substrates of other texture [e.g., (211)] can result in two-fold anisotropy [11]. The main difference in $M(H)$ between the two cases is in the presence of two characteristic steps in $M$ vs. $H$ when the anisotropy is four-fold and only one $M$-vs-$H$ step when it is two-fold. Our VSM and FMR studies of the reference Fe(2 nm) films and Fe/Cr/Fe tri-layers (data not shown) did not reveal any significant in-plane angular dependence in the hysteresis loops or resonance spectra, leading us to conclude that essentially no macroscopic in-plane magnetic anisotropy is present. On the other hand, the numerical analysis described above concludes that the measured one-step-shaped $M(H)$ loops for the RKKY-coupled Fe/Cr/Fe trilayers must be due to two-fold magnetic anisotropy on the scale of the individual crystallites forming the polycrystalline films. The uniform angular distribution of the local anisotropy easy axes in the film plane can result from the deposition on rotating substrates in the case of our samples. Such pattern of magnetic anisotropy can then be explained in terms of a polycrystalline nature of the sputtered multilayers and in-plane strain variations between the nano-crystalline grains [18].

$M(H)$ curves for the model system F1/NM/F2, simulated for different strengths of the AFM interlayer exchange coupling (effective field $H_J$) and shown in Fig. 3(b), exhibit all the key features found in the experimental curves [Fig. 3(a)]. $M(H)$ for Fe/sp1($x$ = 35 at.%)/Fe undergoes a significant change with increasing temperature. The changes are due to the weakening of the interlayer coupling, which can be directly compared the simulated $M(H)$ shown in Fig. 3(b). All of the changes seen in the experimental $M(H)$ data, including the enhancement of the coercivity as the interlayer coupling is weakened, are in correlate very well with the simulated behavior, which validates the model. One should note that the model calculations are performed without taking into account the effect of temperature directly (only via effectively reduced $H_J$), which should reduce magnetic coercivity of the individual layers. This is the likely cause for somewhat smaller coercivity on the experiment.

The simulated $M(H)$ curves shown in Fig. 3(b) are obtained by averaging the $M(H)$ calculated for different angles $\varphi_\mathbf{H}$ between the external field **H** and the easy axis of the uniaxial magnetic anisotropy. Fig. 3(c) shows the curves at selected angles $\varphi_\mathbf{H}$ for the case $H_J/H_a^{av} = 2$. Here, $H_a^{av} = (H_{a1} + H_{a2})/2$, where $H_{a1}$ and $H_{a2}$ are the effective fields of the in-plane uniaxial anisotropy acting in the F1 and F2 layers, respectively. Ratio $H_{a1}/H_{a2} = 0.7$, used in the calculation, corresponds to the value obtained experimentally [Fig. 2(b)]. The step-like shape and coercivity are well-defined for $\varphi_\mathbf{H} < 60°$. As mentioned above, additional VSM and FMR studies of the reference Fe(2 nm) films and Fe/Cr/Fe trilayers did not reveal any significant in-plane angular dependence in the hysteresis loops or the resonance spectra. Since VSM and FMR measure the integral properties of the samples, we conclude that essentially no macroscopic in-plane magnetic anisotropy is present. On the other hand, the observed coercivity can be attributed only to an in-plane magnetic anisotropy. Additionally, the shape of the experimental $M(H)$ curves is closer to the calculated curves obtained by averaging rather than to any individual curve for a select $\varphi_\mathbf{H}$. Therefore, taking into account the polycrystalline nature of our sputtered multilayers, one can conclude that the Fe(2 nm) layers have a uniform angular distribution of the local anisotropy easy axes in the film plane.

Fig. 3(d) illustrates how the energy $U(\varphi_m, \varphi_d)$ of Eq. 2 changes in response to $H$. We, again, take $H_J/H_a^{av} = 2$ and $\varphi_\mathbf{H} = 15°$, which corresponds to the second curve in panel (c). The solid thick line in Fig. 3(d) traces the path connecting the energy minima for different $\varphi_m (\varphi_d)$. The local energy minima are well-defined within this minimum-value path. The minimum at low field corresponds to the antiparallel orientation of the Fe moments ($\varphi_m \approx 90°$, $\varphi_d \approx 180°$). With increasing $H$, a second local energy minimum emerges and deepens, while the first minimum become shallower and eventually disappears. This single-minimum state corresponds to the parallel orientation of the Fe moments ($\varphi_m \approx \varphi_\mathbf{H}$, $\varphi_d \approx 0°$). On subsequently decreasing $H$, the system initially is in the second minimum (parallel magnetic state) until it disappears at lower $H$ and the system ends up in the first energy minimum (antiparallel state).

**Competition between direct and indirect exchange coupling: temperature dependence of magnetic coercivity**

While the first series of trilayers Fe/sp1/Fe exhibits a thermally-induced transition from the low-temperature AFM interlayer coupling into the high-temperature decoupled state, the second series shows a transition from the low-temperature FM to the high-temperature AFM coupling. For the FM-to-AFM thermal transition in the second case, no external magnetic field is required and the magnetization switching is fully reversible – a key advantage for applications.

Using the model validated through the above analysis of the first series of samples, we next focus on investigating the competition between the direct and indirect interlayer exchange coupling in Fe/sp2*($x$)/Fe, with uniform spacers of type sp2* = $Fe_xCr_{100-x}$(1.5 nm) and composite spacers of type sp2* = $Cr(d_{Cr})/Fe_xCr_{100-x}(d)/Cr(d_{Cr})$, $d + d_{Cr} = 1.5$ nm. (sp2* is a derivative from the thickness-fixed spacer sp2 = $Cr(0.4)/Fe_xCr_{100-x}(0.7)/Cr(0.4)$ of the second series). Fig. 4 compares the experimental $M(H)$ loops for the structures with sp2 = $Cr(0.4)/Fe_{15}Cr_{85}(0.7)/Cr(0.4)$ [panel (a)] and the corresponding $M(H)$ curves simulated with $H_J$ chosen such as obtain the best fit to the experiment. First to be noted is the high similarity between the calculated loops and those measured, with all the key features reproduced. Secondly, the experiment shows a temperature-induced transition from the FM interlayer coupling [low-temperature single-loop in Fig. 4(a)] to the AFM coupling [high-temperature loop with zero remanence in Fig. 4(a)]. The variation in the shape of the simulated loops for various effective coupling field values $H_J$ [Fig. 4(b)] additionally confirm the validity of the chosen phenomenological description. Same as in the previous section, $H_{a1}/H_{a2} = 0.7$ was used in the simulations. It should be noted that, even though not the case here, the step-like $M(H)$ shape taken to be due to the AFM interlayer coupling (for example, loops at 300 K and $H_J = 0.5H_a^{av}$), can in principle be caused by different coercive fields in F1 and F2 in the absence of interlayer coupling ($H_J = 0$). Strong FM interlayer coupling, however, always results in a single $M(H)$ loop.

Coercivity of the partial loops ($H_c^{part}$) has a pronounced temperature dependence for all samples and increases almost linearly with decreasing temperature. Fig. 5(a) shows the temperature dependence of the coercive field defined as the difference between the fields of the two peaks on the magnetization derivative, d$M$/d$H$ vs $H$. The series with $x = 15$ % contains samples with different thickness of the layers composing the spacer: $d$ ($d_{Cr}$) = 3 (6), 7 (4), 9 (3), 11 (2), 15 (0) Å. The last sample [$d$ ($d_{Cr}$) = 15 (0) Å] is the trilayer with a uniform spacer $Fe_{15}Cr_{85}$ (1.5 nm). The samples with $d \leq 7$ Å ($d_{Cr} \geq 4$ Å) show a monotonous increase in $H_c^{part}$ with decreasing temperature. The coercivity of the samples with smaller $d_{Cr}$ (< 4 Å) begins to deviate from this slope right below the transition temperature. The high-temperature part of $H_c^{part}(T)$, however, is on the general linear trend [shown as a thick red line in Fig. 5(a)]. This linear slope in the coercive field versus temperature is associated mainly with the change in the intrinsic coercivity of the outer Fe (2 nm) layers.

In our previous work [9] the structures with the spacer thickness of $d \leq 7$ Å ($d_{Cr} \geq 4$ Å) showed the sharpest thermo-magnetic switching. We then suggested that the reason for such narrowing of the magnetic transition was switching off the direct exchange channel between the outer Fe layers. On the other hand, the dependence of $H_c^{part*}$ vs $T$ [Fig. 5(b)], obtained by normalizing $H_c^{part}(T)$ to the sloping intrinsic-coercivity background, shows a noticeable *negative* deviation only for the structures with thin Cr spacers ($d_{Cr} < 4$ Å) and essentially no deviation for $d_{Cr} \geq 4$ Å. The dependence for $x = 20$ %, $d_{Cr} = 4$ Å is shown for comparison because the transition for $x = 15$ %, $d_{Cr} = 4$ Å ($T_C^* \approx 140$ K) is close to the lowest measurement temperature. The absence of a negative deviation on $H_c^{part*}$ vs $T$ for the structures with $d_{Cr} \geq 4$ Å can serve as an addition confirmation that the direct interlayer coupling is fully suppressed.

To separate and analyze the part of the dependence $H_c^{part}(T)$, which is driven by changes in the strength and sign of the interlayer coupling ($H_J$), the coercivity of the simulated $M(H)$ is plotted versus $H_J$ in Fig. 5(c). Thus obtained $H_c^{sim}$ vs $T$ depends on the ratio between the effective anisotropy fields of the F1 and F2 layers, $H_{a1}/H_{a2}$. The larger the deviation of $H_{a1}/H_{a2}$ from unity, the deeper the minimum and bigger its offset from zero field on the FM side of the diagram ($H_J < 0$). When the anisotropy fields are equal ($H_{a1}/H_{a2} = 1$), the minimum is not present. This behavior is similar to the difference between $H_c^{part*}(T)$ for the structures with uniform and composite spacers with large $d_{Cr}$ ($\geq 4$ nm) [blue and black curves in Fig. 5(b), respectively]. This indicates that these two types of spacers transmit the interlayer coupling between the two outer Fe layers differently. In the uniform spacer, direct FM exchange competes with indirect AFM exchange, at some temperature compensating it such that $H_J = 0$. This case is well described by our model, where the F1 and F2 layers have different anisotropy fields [blue curve in Fig. 5(c)]. In contrast, the Fe layers in the structure with the composite spacer are FM coupled at low temperature sequentially through Fe/Cr/FeCr and FeCr/Cr/Fe, with the spacers' FeCr inner layer is in FM state. Since this FeCr layer acts as an addition exchange link, the spacer transmits exchange in such a way as to effectively equalize the coercivity of the outer Fe layers [black curve in Fig. 5(c)]. When the FeCr layer is in its paramagnetic state, the system behaves similar to the one with the uniform spacer [high temperature part of the $H_c^{part*}$ vs $T$ dependence in Fig. 5(b) and the AFM side of $H_c^{sim}$ vs $T$ ($H_J > 0$) in Fig. (c)].

**Conclusions**

In summary, we have described and compared two mechanisms of temperature-induced magnetization switching in multilayers with different types of interlayer exchange mediating spacers. The switching mechanisms reflect the competition of either the direct and indirect exchange coupling through a uniform spacer or the all-indirect exchange coupling of ferromagnetic and antiferromagnetic types through a composite spacer. The key element of the spacer design is the weakly-magnetic diluted-alloy layer, the Curie transition of which is transformed into a P-AP magnetization switching in the structure. Our measured data, supported by detailed theoretical simulations of the magnetic hysteresis in the multilayer, are explained as due to nanograins of uniaxial magnetic anisotropy with its easy axes uniformly distributed in the plane of the outer ferromagnetic layers. The temperature dependence of the magnetic coercivity in the magnetic transition region has a different form for different spacer designs. The specific behavior for the structure with the composite spacer is found to be a result of the suppressed direct interlayer exchange channel, such that the relevant P-AP switching mechanism is a competition of indirect ferromagnetic and indirect antiferromagnetic (both RKKY type) exchange.

We thus have shown that the broken channel of direct interlayer exchange within the spacer is correlated with the sharper thermo-magnetic transition. We furthermore have shown that that the thermally driven competition of the purely indirect interlayer exchange, ferromagnetic RKKY versus antiferromagnetic RKKY, where the proximity effect in the spacer is out of action, leads to even better switching performance. These results should be important for device applications of the Curie-RKKY nanostructures.

**List of abbreviations**

IEC: Indirect exchange coupling; RKKY: Ruderman-Kittel-Kasuya-Yosida; FM: Ferromagnetic; PM: Paramagnetic; NM: Nonmagnetic; AFM: Antiferromagnetic; P: Parallel; AP: Antiparallel; VSM: Vibrating-sample magnetometer; MOKE: Magneto-optical Kerr effect; FMR: Ferromagnetic resonance.

**Declarations**

**Availability of data and material**

All data are fully available without restriction.

**Competing interest**

The authors declare that they have no competing interests.

**Funding**

This work was partly financially supported by the Swedish Stiftelse Olle Engkvist Byggmästare (Grant No. 2013-424/2018-589), the Swedish Research Council (VR Grant No. 2014-4548), the Department of Targeted Training of Taras Shevchenko National University of Kyiv at the National Academy of Sciences of Ukraine (Grant No. 0117U006356), Volkswagen Foundation (Grant No. 90418), and the State Fund for Fundamental Research of Ukraine (Grant No. F76/63-2017).

**Authors' contributions**

DP, YTP, AK and VK developed the approaches to fabricate multilayers and carried out main works on characterization of the samples. DP and YTP carried out temperature dependent magnetic measurements on multilayers. DP, VB and AT analyzed the data of magnetic measurements and made calculations to extract magnetic parameters of the multilayers. All authors contributed to analysis of experimental data and writing manuscript. AK, AT and VK supervised the work and finalized the manuscript. All authors read and approved the final manuscript.

All authors have the appropriate permissions and rights to the reported data. All authors have agreed to authorship and order of authorship.

**Acknowledgements**

Not applicable.

**Authors' information**

[1]Nanostructure Physics, Royal Institute of Technology, 10691 Stockholm, Sweden. [2]Institute of Magnetism, NAS of Ukraine and MES of Ukraine, 03142 Kyiv, Ukraine.

**Figures captions**

Fig. 1. Illustration of magnetic layout of Fe/uniform-spacer/Fe multilayers when the spacer is paramagnetic (PM) (a) or ferromagnetic (FM) (b). (c), (e) Structures with modified, composite spacers sp1 and sp2 exhibit, respectively, antiparallel and parallel magnetic ground state at low temperature ($T < T_C^*$). (d), (f) Corresponding characteristic temperature variation of remanent magnetization of structures with spacers sp1 and sp2 for different compositions of spacers' inner diluted alloy layer. Layer thicknesses are given in parentheses in "nm".

Fig. 2. (a) Typical in-plane magnetization curve, $M(H)$, measured by MOKE for Fe/sp2/Fe multilayers with antiferromagnetic interlayer coupling. Curved arrows show direction of field sweep; horizontal arrows denote mutual alignment of Fe magnetic moments. (b) MOKE $M(H)$ loops for reference Fe(2)/Cr(10) (bottom Fe) and Cr(10)/Fe(2) (top Fe) bilayers. (c) Reference-frame schematic of in-plane **M**$_1$, **M**$_2$, and **H**, with respect to easy axis of two-fold magnetic anisotropy of a nano-crystallite.

Fig. 3. (a) Measured $M(H)$ curves for a sample from series, Fe/sp1($x$ = 15 %)/Fe, for different temperatures. (b) Corresponding simulated $M(H)$ curves for model F1/NM/F2 trilayer for different strengths of effective field $H_J$ of indirect exchange coupling. ($H_a^{av} = (H_{a1} + H_{a2})/2$, where $H_{a1}$ and $H_{a2}$ are anisotropy fields of layers F1 and F2. (c) $M(H)$ curves simulated for selected angles $\varphi_H$, for $H_J/H_a^{av}$ = 2. (d) Transformation of local minima of free energy (2) as a function of applied field $H$, for case $H_J/H_a^{av}$ = 2 and $\varphi_H$ = 15º. Blue lines trace the path connecting energy minima for different $\varphi_m$ ($\varphi_d$). Front surface of energy surface is transparent for visual clarity of illustration.

Fig. 4. (a) Magnetization versus field measured by the MOKE for sample from second series, Fe/sp2($x$ = 15 %)/Fe, for different temperatures. (b) Corresponding simulated $M(H)$ curves for model F1/NM/F2 trilayer, for different effective field of indirect exchange coupling, $H_J$.

Fig. 5. (a) Temperature dependence of coercivity of partial loops ($H_c^{part}$) for structures Fe/sp2($x$ = 15 %)/Fe with different thickness of Fe$_x$Cr$_{100-x}$ and Cr layers ($d$ and $d_{Cr}$, respectively) in spacer sp2. Red thick line is linear approximation of high-temperature part of $H_c^{part}(T)$. (b) Temperature dependence of coercivity normalized to linear background. (c) Coercivity vs. $H_J$ obtained from simulated $M(H)$ curves for two cases: (1) $H_{a1}/H_{a2}$ = 0.7 and (2) $H_{a1} = H_{a2}$.

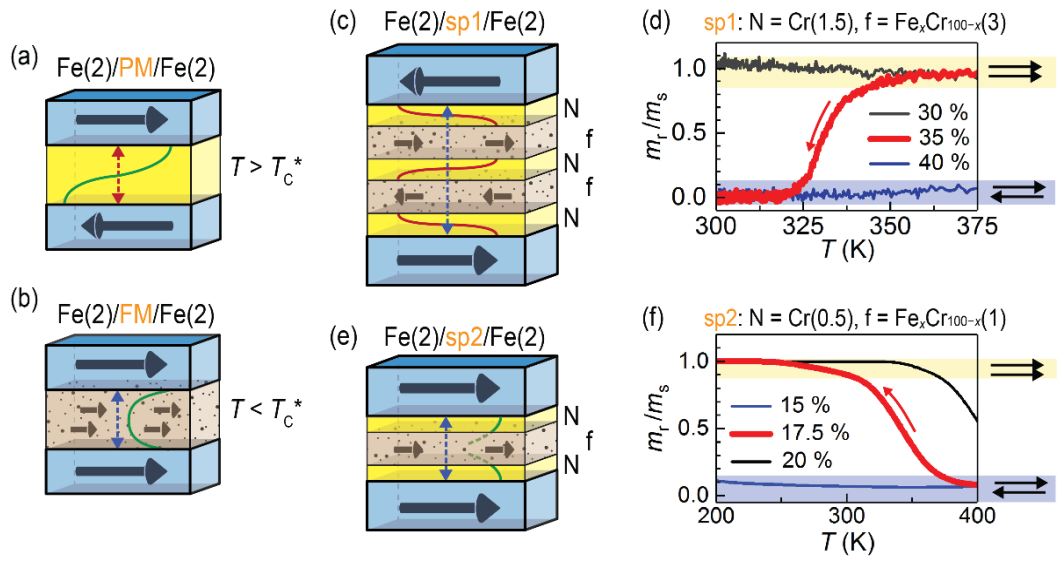

Fig. 1

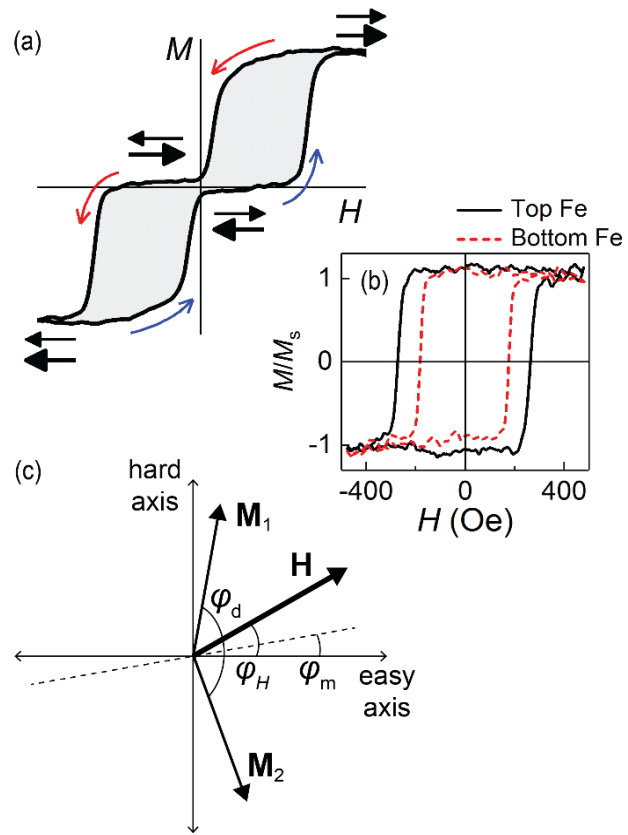

Fig. 2.

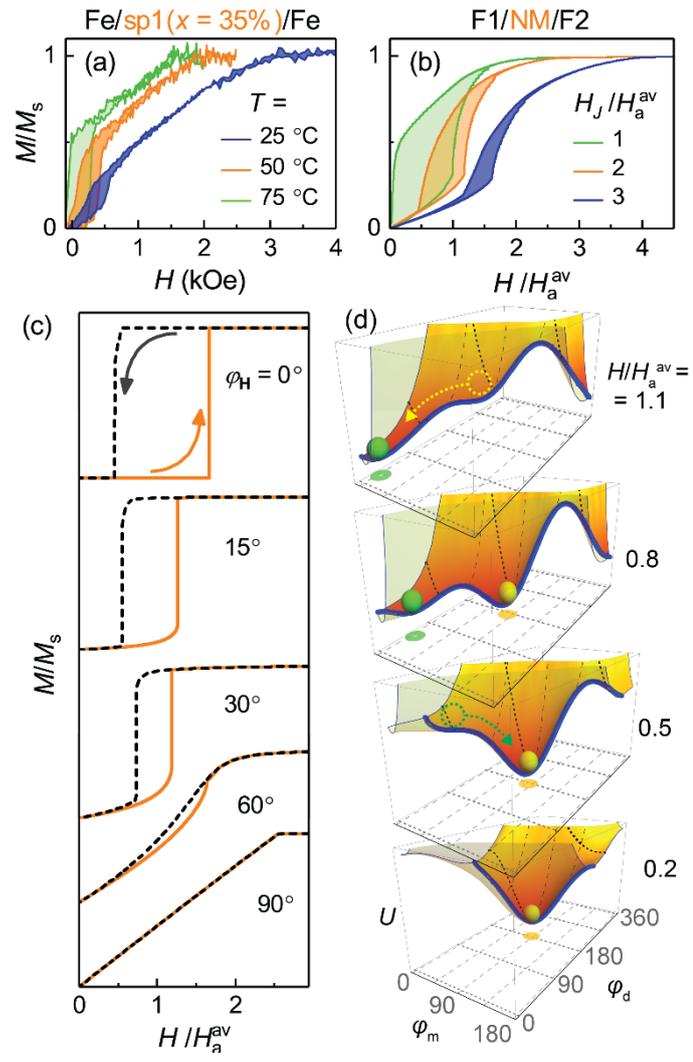

Fig.3.

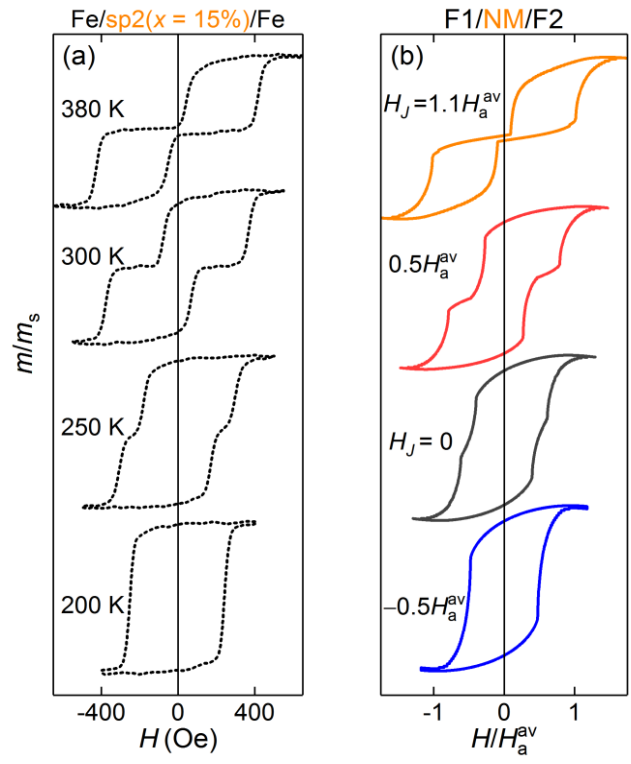

Fig4.

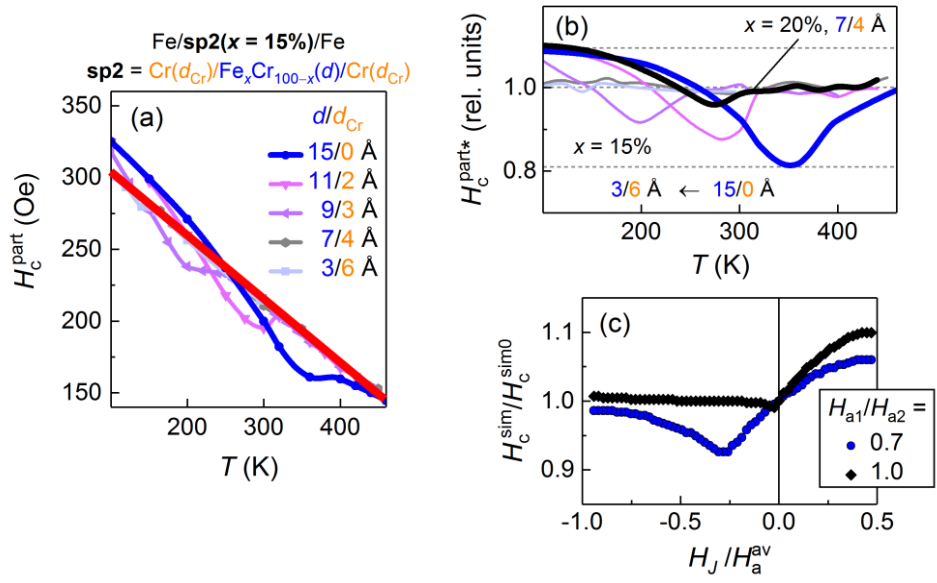

Fig.5.